\documentclass[amsmath,amssymb,aps,prb,twocolumn,showpacs,floatfix]{revtex4}
\usepackage{graphicx}
\usepackage{dcolumn}
\usepackage{bm}
\usepackage{color}
\newcommand{\trace}{{\rm Tr}}

\begin{document}

\title{On the nature of heat in strongly coupled open quantum systems}

\author{Massimiliano Esposito}
\affiliation{Complex Systems and Statistical Mechanics, Physics and Materials Science Research Unit, 
University of Luxembourg, L-1511 Luxembourg, Luxembourg}
\author{Maicol A. Ochoa}
\affiliation{Department of Chemistry \& Biochemistry, University of California San Diego, La Jolla CA 92093, USA}
\author{Michael Galperin}
\affiliation{Department of Chemistry \& Biochemistry, University of California San Diego, La Jolla CA 92093, USA}

\date{\today}

\begin{abstract}
We study heat transfers in a single level quantum dot strongly coupled to fermionic reservoirs and subjected 
to a time-dependent protocol modulating the dot energy as well as the dot-reservoir coupling strength. 
The dynamics is described using nonequilibrium Greens functions (NEGFs) evaluated to first order beyond quasi-static driving. 
We show that any heat definition expressed as an energy change in the reservoir energy plus any fraction of 
the system-reservoir interaction is not an exact differential when evaluated along reversible isothermal 
transformations, except when that fraction is zero. However, even in that latter case the reversible heat 
divided by temperature, namely the entropy, does not satisfy the third law of thermodynamics and diverges 
in the low temperature limit. Our results cast doubts on the possibility to define a thermodynamically 
consistent notion of heat expressed as the expectation value of some Hamiltonian terms.
\end{abstract}


\maketitle


The nature of heat is one of the most fundamental questions which has been driving research in 
thermodynamics since its origins. Nowadays, establishing a thermodynamically consistent notion 
of heat for open quantum system is of crucial importance for mesoscopic physics and for the study 
of energy conversion in small devices. This issue has direct implications on defining meaningful 
notions of efficiency in thermoelectricity or photoelectricity for instance. 

For systems weakly interacting with their reservoirs the situation is rather clear 
\cite{SpohnLebowitz78, Breuer02, Esposito12, Kosloff13, Kurizki15, EspoBulnEng15, EspositoReview}. 
The heat flux is defined as minus the energy change in the reservoir and can be directly related to 
the system energy changes since the system-reservoir coupling energy is negligible.  
This definition has been extensively used to study the performance of a broad range of nano-devices 
(see e.g. \cite{EspoLindVdB_EPL09_Dot, EspoRuttCleuPRB, EspoKawLindVdB_PRE_10, EspoKumLindVdBPRE12, 
ButtikerSanchez10PRL, ButtikerSanchez11PRL, EntinWohlmanAharonyPRB12, EntinWohlmanImryAharonyPRB10, 
BenentiCasatiRev13, LutzSingerPRL14, AlonsoCorreaPRE13, UzdinKosloff15, EspoGalpPRB15}). 

The situation is also clear in the strong coupling regime, as long as the system 
operates in a  steady state \cite{WhitneyPRB13, GaspardNJP15, BrandesSchaller15EPL} 
(see also e.g. \cite{HanggiSaitoDharPRE12, PazPRL13}). 
Indeed attributing the coupling energy to the system or to the reservoirs is equivalent 
in this case since net changes in the coupling energy are zero. The first law reduces to 
Kirchhoff's law for heat fluxes crossing the system and the second law reduces to the non-negativity 
of $-\sum_\nu \dot{Q}_{\nu}/T_{\nu} \geq 0$ where $\dot{Q}_{\nu}$ is the heat entering 
the system from reservoir $\nu$ and $T_{\nu}$ is the temperature of that reservoir.
This result can easily be shown using scattering theory or nonequilibrium Green's functions (NEGF) approaches. 
Many performance studies have thus considered steady state setups (see e.g. \cite{LinkePRL02, LinkePRL05, 
EspoSchallerNJP13, EspoKrauseJCP15, WhitneyPRL14, SeifertBrandnerPRL13, SeifertBrandnerNJP13}).

However, the situation is very different when considering setups where the system is driven by a 
time-dependent processe since in this case the changes in the coupling energy must be accounted for. 
Despite the fact that such setups are indispensable to consider reversible transformations which 
play a central role in thermodynamics, few studies have considered them because the dynamics 
typically becomes difficulty to solve. 
We recently proposed a consistent nonequilibrium thermodynamics formulation for noninteracting 
quantum systems strongly coupled to their reservoirs and driven by a slowly changing external 
field \cite{EspGalpPRL15}. 
Within the framework of NEGF we calculated transport characteristics 
to first order beyond quasi-static limit.
This formulation has the particularity that not only heat but all the other thermodynamic 
quantities such as work, system energy and entropy have no simple expression in terms of 
quantum expectations values of operators.   
In this letter, we use the same framework of NEGF to show that any attempt to define heat 
in term of quantum expectations values of operators leads to thermodynamic inconsistencies.

The typical Hamiltonian of an open quantum system $S$ coupled to multiple reservoirs $\nu$ 
at temperatures $T_\nu$ and chemical potentials $\mu_\nu$ is 
\begin{equation}
\hat H(t)=\hat H_S(t) + \sum_\nu \left( \hat H_\nu + \hat V_\nu(t)\right), \label{TotHam}
\end{equation}
where $\hat H_S$ ($\hat H_\nu$) denotes the system (reservoir $\nu$) Hamiltonian 
and $\hat V_\nu$ is the system-reservoir interaction.

We start by introducing the class of all possible heat definitions expressed as the 
change in the quantum expectation value of the reservoir Hamiltonian plus a fraction 
$0 \leq \alpha \leq 1$ of the the system-reservoir coupling energy 
(we set $\hbar=e=k_B=1$ throughout the paper)
\begin{align} 
\label{TradHeat}
\dot{Q}_{\nu,\alpha} = J_{\nu,\alpha} - \mu_\nu I_{\nu},
\end{align}
where the matter and heat currents entering the system from reservoir $\nu$ are given by
\begin{align} \label{defI}
I_{\nu} =& -\trace \{\hat N_{\nu}\, d_t \hat{\rho}\}
= -d_t \langle \hat N_\nu \rangle \\ \label{defJ}
J_{\nu,\alpha}= & -\trace \{\big( \hat H_{\nu}+\alpha \hat V_{\nu} \big) d_t \hat{\rho}\},
\end{align}
and $\hat\rho(t)$ is the density matrix of the total system.
The heat flux definition most commonly used in the literature corresponds to the choice 
$\alpha=0$ and can be expressed in terms of the rate of change in the number operator 
$\hat N_{\nu}$ and in the Hamiltonian $\hat H_{\nu}$ of the reservoir $\nu$, 
since $J_{\nu,0}=-d_t \langle \hat H_\nu \rangle$
\cite{MahanBook, GalperinNitzanRatner_heat_PRB07, WuSegalJPhysA09, SegalPRB13, 
EspoLindVdBNJP10, EspoPucciPeliti13, PazPRL13}.
The choice $\alpha=1$ was considered for instance in Ref.~\cite{AllahverdyanPRE01}
and the choice $\alpha=1/2$ in Ref.~\cite{SanchezPRB14}. 

The specific model that we will consider consists of an externally driven level $\varepsilon(t)$
bi-linearly coupled to a single Fermionic reservoir $R$ at equilibrium. 
Its Hamiltonian is 
given by (\ref{TotHam}), where the level, the reservoir and their coupling respectively read 
\begin{align}\label{ModelH}
&\hat H_S(t) = \varepsilon(t) \hat d^\dagger\hat d \ \ , \ \ \hat H_R = \sum_{k} \varepsilon_k c_k^\dagger\hat c_k \\
\label{ModelV}
& \hat V(t) = \sum_{k}\left( V_{k}(t)\hat d^\dagger\hat c_k+\mbox{H.c.}\right).
\end{align}
Here $\hat d^\dagger$ ($\hat d$) and $\hat c_k^\dagger$ ($\hat c_k$) create (annihilate) 
an electron in the level of the system and in state $k$ of the reservoir, respectively. 
$\varepsilon_k$ is the energy of the latter.
We emphasize that the external driving can modify the position of the level, $\varepsilon(t)$, 
as well as the strength of the system-reservoir coupling, $V_{k}(t)$.
Following Ref.~\cite{JauhoWingreenMeirPRB94}, we assume that this latter is of the form
\begin{equation}
\label{ut}
V_{k}(t) = u(t)\, V_{k}\qquad u(t)\in \mathbb{R}.
\end{equation}
For the simulations presented in this letter we will consider the driving protocols 
\begin{align}
\label{et}
\varepsilon(t)=&\varepsilon_0+\frac{\Delta_\varepsilon}{2}\bigg(1-\cos\omega_0 t\bigg)
\\
\label{gt}
u^2(t)=&1+\frac{\Delta_\Gamma}{2\Gamma_0} \bigg(1-\cos\omega_0 t\bigg).
\end{align} 

The explicit expression of the heat flux (\ref{TradHeat}) in terms of NEGF 
can be found in Eqs. (S7)-(S10) of the supplementary material \cite{SupMat}. 
In general a NEGF depends on two times, $t_1$ and $t_2$, but only depends on 
their difference $\tau=t_1-t_2$ at steady state. If the driving acting on the 
system is slow compared to the system relaxation timescale, after a Fourier 
transform in $\tau \to E$, one can make use of the slow time-dependence of 
the resulting NEGF in $t=(t_1+t_2)/2$ to evaluate its equation of motion. 
This procedure is known as the gradient expansion and is detailed in the 
supplementary material \cite{SupMat}.
When using it to evaluate the heat flux for our model (\ref{ModelH})-(\ref{ut}), 
we obtain to the lowest order corresponding to the quasi-static limit  
\begin{equation}
\begin{split}\label{QQS}
&\dot{Q}_{\alpha}^{(1)} =
\frac{d}{dt}\bigg( \int\frac{dE}{2\pi}f\, A^{(0)}\big[(E-\mu) +(1-2\alpha)(E-\varepsilon) \big]\bigg) \\
&-\int\frac{dE}{2\pi}f \bigg( A^{(0)}\, d_t{\varepsilon}+(1-\alpha)\bigg[\mbox{Re}\, G^{r(0)}\, 
\partial_t{\Gamma}+A^{(0)}\, \partial_t \Lambda\bigg] \bigg),
\end{split}
\end{equation}
where $f(E)=[e^{(E-\mu)/T}+1]^{-1}$ is the Fermi-Dirac distribution in the reservoir,
the zero order retarded Green function is given by
\begin{align}
\label{grad1Gr}
G^{r(0)}(t,E)=&\left[E-\varepsilon(t)-\Lambda(t,E)+i\Gamma(t,E)/2 \right]^{-1}
\end{align}
and $A^{(0)}(t,E)=-2\, \mbox{Im}\, G^{r(0)}(t,E)$ is the system spectral function. 
The Lamb shift and broadening caused by coupling to the reservoir are taken as\cite{WingreenMeirPRB94,MahanBook}
\begin{align}
\label{defLambda}
\Lambda(t,E) =& u^2(t) \, \Gamma_0\,\frac{1}{2}\frac{(E-E_B)W_B}{(E-E_B)^2+W_B^2} \\
\label{defGamma}
\Gamma(t,E) =& u^2(t) \, \Gamma_0\,\frac{W_B^2}{(E-E_B)^2+W_B^2},
\end{align}
where $E_B$ and $W_B$ are the center and width of the band, respectively.
To our knowledge (\ref{QQS}) is the first explicit expression for a quasi-static heat of the kind (\ref{TradHeat}).
Two major results ensue. 

\begin{figure}[htbp]
\vspace{-1cm}
\centering\rotatebox{0}{\scalebox{0.4}{\includegraphics{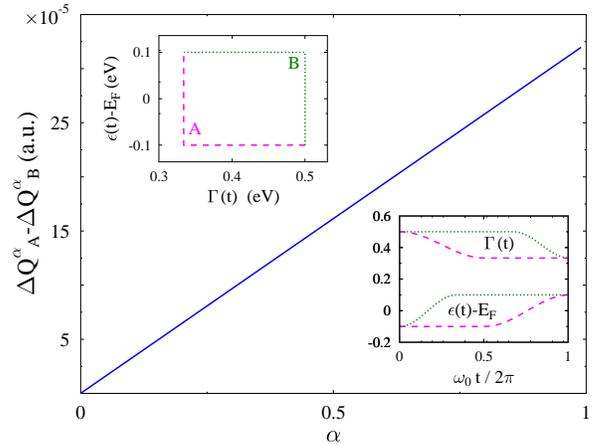}}}
\vspace{-1cm}
\caption{\label{nfig1}(Color online) 
Difference between the quasi-static heat produced along two different driving protocols 
denoted by A and B and corresponding to (\ref{et}) and (\ref{gt}) with parameters 
$T=10$~K, $\varepsilon_0=-0.1$~eV, $\Delta_\varepsilon=0.2$~eV, 
$\Delta_\Gamma=-1/6$~eV, $\Gamma_0=0.5$~eV, $\omega_0=10^{13}$~s${}^{-1}$. 
The band parameters are $E_B=0$ and $W_B=5$~eV and the Fermi energy is $E_F=0$.
The two protocols are shown in the left top inset and the time dependence of the level position 
and coupling strength corresponding to the protocols are given in the bottom right inset.
}
\end{figure}

A central requirement in thermodynamics is that the reversible heat change is an exact differential. 
This implies that mixed derivatives of the heat rate with respect to the driving parameters 
$\varepsilon(t)$ and $u(t)$ should be equal to each other
\begin{equation}
\label{drv2}
 \frac{\partial^2 Q^{(1)}_\alpha}{\partial\varepsilon\,\partial u}
 =  \frac{\partial^2 Q^{(1)}_\alpha}{\partial u\,\partial\varepsilon}.
\end{equation}
Our first important result is that this property is only satisfied for $\alpha=0$. 
For any other choice of $\alpha$, the reversible heat is not an exact differential 
and thus cannot be considered as a thermodynamically consistent definition.  
This result can be explicitly seen in Fig.\ref{nfig1} where two different reversible 
driving protocols connecting the same initial and final point give rise to different 
reversible heat except for $\alpha=0$. 

\begin{figure}[htbp]
\vspace{0cm}
\centering\includegraphics[width=\linewidth]{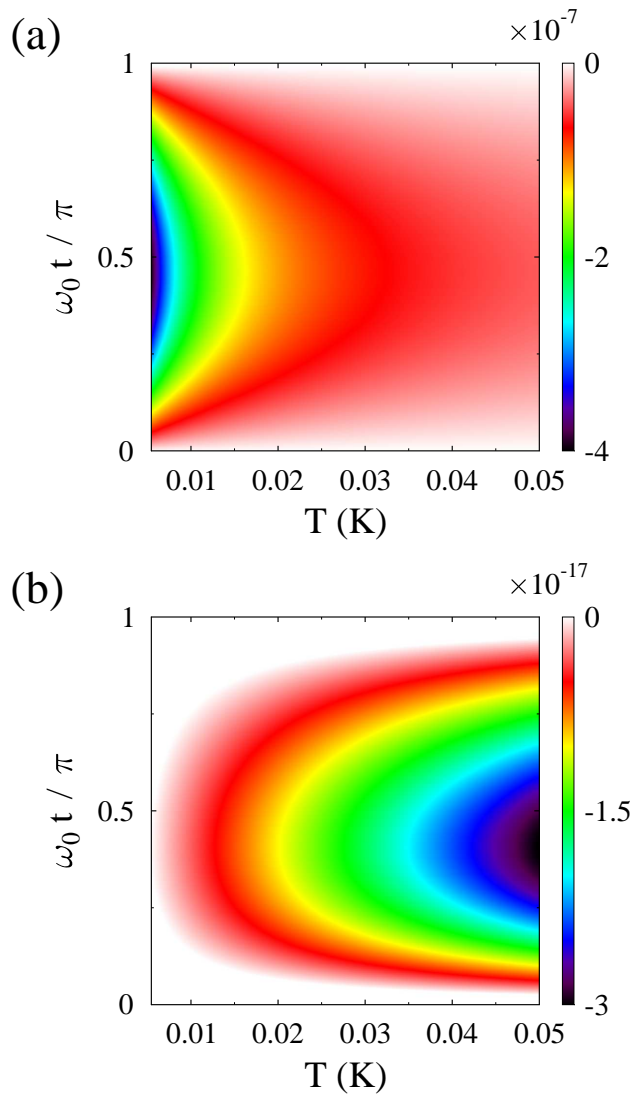}
\vspace{-0.5cm}
\caption{\label{nfig2}
(Color online) Entropy change ($J/K\, s$) given by (a) Eq.(\ref{EqEntropy}) and (b) Eq.(14) 
in Ref.~\cite{EspGalpPRL15}, as a function of temperature and the time along half of the 
period performed by the driving (\ref{et}) and (\ref{gt}) with 
$\varepsilon_0=\Delta_\varepsilon=0$, $\Delta_\Gamma=1/6$~eV, $\Gamma_0=0.5$~eV,
$\omega_0=10^{14}$~s${}^{-1}$. The other parameters are the same as in Fig.~\ref{nfig1}.
}
\end{figure}

Our second important result is that since the equilibrium entropy is the state function whose 
differential is the reversible heat divided by temperature $d_t S^{eq}=\dot{Q}_{0}^{(1)}/T$, 
by integrating the reversible heat rate (\ref{QQS}), we are able to find the equilibrium 
entropy up to a constant (see \cite{SupMat}) 
\begin{align}
S^{eq}=& \int\frac{dE}{2\pi}A \big(-f\ln f-[1-f]\ln [1-f]\big) \nonumber \\
&+ \int\frac{dE}{2\pi} A f \frac{(E-\varepsilon)}{T} \label{EqEntropy} \\
&+ \int\frac{dE}{2\pi} A \ln [1-f] \big(\partial_E \Lambda
 +\frac{E-\varepsilon-\Lambda}{\Gamma} \partial_E \Gamma \big). \nonumber
\end{align}
The first contribution has the appealing form of an energy resolved equilibrium entropy.
The second one is exactly half of the equilibrium expectation value of the coupling 
energy divided by temperature, namely $\langle \hat V_{\nu}(t)\rangle^{eq}/(2T)$. 
The third one is due to the energy resolution of the Lamb shift and broadening and thus 
vanishes in the wide-band limit when $\Lambda\to 0$ and $\Gamma$ does not depend on energy.
In the low temperature limit $T \to 0$, the first terms goes to zero as expected by the 
third law of thermodynamics, but the other two terms diverge, casting doubts
on the thermodynamic relevance of the heat definition $\dot{Q}_{0}$. 
The only way to avoid the divergence is to send the coupling strength to zero before 
taking the low temperature limit. Indeed, in this case the first term becomes the weak 
coupling Shannon entropy and the last two vanish. While one may have expected that the 
finite coupling can create a finite entropy in the system at low temperature, justifying 
a divergent entropy is more difficult and seems pathological.    
In Figure \ref{nfig2} we compare the behavior of the entropy change obtained from the 
reversible heat $Q_{0}^{(1)}$ given by (\ref{QQS}) and the reversible heat ${\cal Q}$ 
that we recently proposed in Ref. \cite{EspGalpPRL15}. The low temperature divergence 
is clearly seen in the first case but not in the second one, as proved in \cite{EspGalpPRL15}. 

We now consider the heat generated along the cycle of a periodic driving when 
the system reached a stationary regime (i.e when initial transients are gone).
Since the quasi-static heat vanishes along a cycle, on must calculate its 
second order contribution. Its general expression is derived in the 
supplementary material \cite{SupMat}. When integrated over a cycle of 
duration $\tau$ for $\alpha=0$, the resulting heat reads $Q_{0}^{(2)}=$
\begin{equation}
\int_0^{\tau} dt \int\frac{dE}{2\pi}\partial_E f \frac{\left[A^{(0)}\right]^2}{2}
\bigg(d_t\varepsilon+\partial_t\Lambda+\partial_t\Gamma\frac{E-\varepsilon-\Lambda}{\Gamma}\bigg)^2.
\end{equation}
Since $\partial_E f$ is always negative, this heat is always negative 
in agreement with the second law of thermodynamics.

We finally comment on the heat definition $Q_{1/2}$ proposed in Ref.~\cite{SanchezPRB14} 
when considering a strongly coupled {\em ac}-driven resonant level coupled to a single 
reservoir treated by scattering and Floquet theories. The couplings to the reservoirs 
were assumed time-independent ($u$ constant) and the wide band approximation was used. 
We show in the supplementary material that in this limit our treatment reproduces 
the expression for the heat $Q_{1/2}$ found in Ref.~\cite{SanchezPRB14}.
By integrating its quasi-static form, since $T\, d_t S^{eq}_{1/2} = \dot{Q}_{1/2}^{(1)}$, 
we further show that its corresponding equilibrium entropy is given by
\begin{equation}\label{EqEntropySanchez}
S^{eq}_{1/2} = \int\frac{dE}{2\pi}A \big(-f\ln f-[1-f]\ln [1-f]\big).
\end{equation}
This is the first contribution to the entropy found in (\ref{EqEntropy}). 
The second contribution dropped due to the choice $\alpha=1/2$ and the third 
due to the wide-band approximation. Since no driving in the coupling was 
considered, the reversible heat $Q_{1/2}^{(1)}$ is also a state function.
We thus confirm that under the assumptions made in \cite{SanchezPRB14} (wide band 
approximation and no driving in the coupling) the heat definition $\dot{Q}_{1/2}$ 
can be considered as very appealing. However, we proved that this definition fails 
when these assumptions are released.  

We contributed to the fundamental question of the nature of heat in open quantum system 
strongly interacting with a reservoir and driven by a time-dependent force in the system 
and in the system-reservoir energy, within the framework of NEGF.
Our central finding is that any heat definition expressed as the change in the quantum 
expectation value of the reservoir energy plus any fraction $\alpha$ of the coupling 
energy displays thermodynamic inconsistencies. Any $\alpha$ different from zero leads to 
a quasi-static heat which is not a state function. The choice $\alpha=0$ is more appealing 
(the quasi-static heat is a state function and the second law is satisfied for our model)
but leads to an entropy which diverges in the low temperature limit.    
Our considerations were made possible by using the gradient expansion of NEGF
which provides to our knowledge the first explicit quasi-static expression for 
the various heat definitions that we considered. 
This only assumption used in this approach is that the reservoir Greens functions are always thermal.
Our conclusion reinforces our proposal in Ref. \cite{EspGalpPRL15} to abandon heat definitions 
(and other thermodynamic quantities) expressed as quantum expectation values of operators 
in order to derive a consistent thermodynamics within the framework of NEGF for open 
quantum system beyond the weak coupling limit.

\appendix
\section{Particle and energy fluxes}\label{app_fluxes}
We consider the standard definition for the particle and energy fluxes at the interface with 
reservoir $\nu$, Eqs.~(\ref{defI}) and (\ref{defJ}), respectively.
In terms of Green functions, these definitions yield~\cite{JauhoWingreenMeirPRB94,GalperinNitzanRatner_heat_PRB07} 
\begin{align}
\label{InuGF}
 I_\nu(t) =& 2\,\mbox{Re}\,\mbox{Tr}\int dt_1\big\{
 G^{<}(t,t_1)\,\Sigma_\nu^a(t_1,t)
 \\ & \qquad\qquad\quad
 +G^r(t,t_1)\,\Sigma_\nu^<(t_1,t)
 \big\}
 \nonumber \\
 \label{JnuGF}
 J_{\nu,\alpha}(t) =& (\alpha-1)\,\partial_t\langle\hat V_\nu(t)\rangle
 -\alpha\, d_t\langle\hat V_\nu(t)\rangle
 \nonumber \\ +&
 2\,\mbox{Im}\,\mbox{Tr}\int dt_1\big\{
 G^<(t,t_1)\,\partial_t\Sigma_\nu^a(t_1,t)
 \\ & \qquad\qquad\quad
 +G^r(t,t_1)\,\partial_t\Sigma_\nu^<(t_1,t)
 \big\}, \nonumber
\end{align}
where
\begin{align}
\label{VnuGF}
 \langle \hat V_\nu(t)\rangle =& 2\,\mbox{Im}\,\mbox{Tr}\int dt_1\big\{
 G^<(t,t_1)\,\Sigma_\nu^a(t_1,t)
 \\ &\qquad\qquad\quad +G^r(t,t_1)\,\Sigma_\nu^<(t_1,t)
 \big\}. \nonumber
\end{align}
The partial derivatives in the first and third terms in the right side of Eq.(\ref{JnuGF}) 
indicate a time derivative of the system-reservoir coupling only in the external driving.
$\mbox{Tr}\{\ldots\}$ denotes a trace over the system subspace.
$G^{<}=G^{-+}$ and $G^{r}=G^{--}-G^{-+}$  are matrices in the system subspace and 
are the lesser and retarded projections of the single-particle Green function 
\begin{equation}
\label{defG}
G_{mm'}(\tau,\tau') = -i\left\langle T_c\,\hat d_m(\tau)\,\hat d_{m'}^\dagger(\tau') \right \rangle,
\end{equation}
where $T_c$ denotes the contour ordering operator, $\tau$ and $\tau'$ are the contour variables,
and the contour branches are labeled as time ordered, $s={}-{}$, and anti-time ordered, $s={}+{}$.
$\Sigma_\nu^{<}=\Sigma_\nu^{-+}$  and $\Sigma_\nu^{a}=\Sigma_\nu^{-+}-\Sigma_\nu^{++}$ 
are also matrices in the system space and are the lesser and advanced projections 
of the self-energy due to the coupling to reservoir $\nu$
\begin{equation}
\label{defSigma}
\left[\Sigma_\nu(\tau,\tau')\right]_{mm'} = \sum_{k\in\nu} V_{mk}(t)\,g_k(\tau,\tau')\, V_{km'}(t'),
\end{equation}
where
\begin{equation}
g_k(\tau,\tau')\equiv-i\langle T_c\,\hat c_k(\tau)\,\hat c_k^\dagger(\tau')\rangle
\end{equation}
is the equilibrium Green function for the free electrons in the reservoir $\nu$. 
The equations of motion for the projection $s_1 s_2$ of the GF (\ref{defG}) are given by 
\begin{align}
 \label{LEOM}
 &\left(i\frac{\overset{\rightarrow}{\partial}}{\partial t_1}\sigma^z_{s_1s_2}-H_S(t_1)\right)G^{s_1s_2}(t_1,t_2)
 = 
 \\ &\quad
 \sigma^z_{s_1s_2}\delta(t_1-t_2) - \sum_{s_3}\int dt_3\,\Sigma^{s_1s_3}(t_1,t_3)\,
 s_3\, G^{s_3s_2}(t_3,t_2)
\nonumber \\
 \label{REOM}
 &G^{s_1s_2}(t_1,t_2)\left(-i\frac{\overset{\leftarrow}{\partial}}{\partial t_2}\sigma^z_{s_1s_2}-H_S(t_2)\right)
 = 
 \\ &\quad
 \sigma^z_{s_1s_2}\delta(t_1-t_2) - \sum_{s_3}\int dt_3\, G^{s_1s_3}(t_1,t_3)\, 
 s_3\, \Sigma^{s_3s_2}(t_3,t_2),
 \nonumber
\end{align}
where $\mathbf{\sigma}^z$ is the Pauli matrix, and $\Sigma^{s_1s_2}(t_1,t_2)$ is the total 
self-energy, i.e. the self-energy due to the system-reservoirs couplings and the intra-system 
interactions.

\section{Gradient expansion}\label{app_grad}
Green functions and self-energies are two-time functions, $F(t_1,t_2)$. 
Introducing via a change of variable the classical timescale, $t=(t_1+t_2)/2$, 
and the quantum timescale, $s=t_1-t_2$, and performing a Fourier transform in 
the quantum time leads to the time-dependent energy resolved function 
$F(t,E)=\int ds e^{iEs} F(t,s)$, which is the Wigner transform of $F(t,s)$. 
Naturally
\begin{equation}
F(t_1,t_2)=F(t,s)= \int\frac{dE}{2\pi}\, e^{-iE s}F(t,E).
\end{equation}
Below, we will consider partial derivatives of the form $\partial_{t_2}F(t_1,t_2)$
(see Eq.~(\ref{JnuGF})). Their Wigner transforms read $[\partial_t/2+i E] F(t,E)$.
We will also consider integral expression such as
\begin{equation}
 F(t_1,t_2) = \int dt_3\, F_1(t_1,t_3)\, F_2(t_3,t_2),
\end{equation}
whose Wigner transform reads \cite{HaugJauho_2008}
\begin{equation}
 \label{gradtot}
 F(t,E) = F_1(t,E) \, \hat{\mathcal{G}}(t,E)\, F_2(t,E),
\end{equation}
where
\begin{equation}
 \label{gradop}
 \hat{\mathcal{G}}(t,E)=  \mbox{exp}\left(\frac{1}{2i}\left[
 \overset{\leftarrow}{\partial}_t\,\overset{\rightarrow}{\partial}_E-
 \overset{\leftarrow}{\partial}_E\,\overset{\rightarrow}{\partial}_t
 \right]\right)
\end{equation}
is the gradient operator. 
At steady state the dependence on $t$ vanishes and only the energy resolution $E$ survives.
This means that when the driving is slow relative to the characteristic relaxation timescales 
of the system, we can expand (\ref{gradop}) in Taylor series and truncate the series to the suited level.
Traditionally the gradient expansion goes to the first order, but we will need the second order below
\begin{align}
\label{grad2}
&F(t,E)\approx F_1(t,E)\, F_2(t,E)
\\ &\qquad
+\frac{i}{2}\left\{F_1(t,E);F_2(t,E)\right\}
 - \frac{1}{8}\left[F_1(t,E);F_2(t,E)\right],
 \nonumber
\end{align}
where
\begin{align}
& \left\{F_1(t,E);F_2(t,E)\right\}= 
 \\ &\quad
\partial_E F_1(t,E)\,\partial_t F_2(t,E)-\partial_t F_1(t,E)\,\partial_E F_2(t,E)
\nonumber
\\
& \left[F_1(t,E);F_2(t,E)\right]= \partial^2_E F_1(t,E)\,\partial^2_t F_2(t,E)
 \\ &\quad
 +\partial^2_t F_1(t,E)\,\partial^2_E F_2(t,E)
-\partial^2_{tE} F_1(t,E)\,\partial^2_{tE} F_2(t,E)
\nonumber .
\end{align}

Below we will also need to consider the dependence of the full self-energy 
$\Sigma(t_1,t_2)$ on the system-reservoir coupling $u(t)$. Since
\begin{equation}
 \Sigma(t_1,t_2) = u(t_1)\, S(t_1,t_2)\, u(t_2),
\end{equation}
it is easy to show that up to second order gradient expansion, the functions $\Sigma$ and $S$ are related by
\begin{align}
 \Sigma(t,E)\approx& u^2(t)\, S(t,E) 
 \\ -&
 \frac{1}{4}\left( \partial^2_t u(t) - [\partial_t u(t)]^2 \right)\partial_E^2 S(t,E).
 \nonumber 
\end{align}
Similarly their time derivatives are related by
\begin{equation}
 \partial_t\Sigma(t,E) \approx u^2(t)\,\partial_t S(t,E) + u(t)\,\partial_t u(t)\, S(t,E).
\end{equation}

\section{Slow driving of a single level coupled to a reservoir}\label{app_onelevel}
We now restrict our consideration to a single level, Eqs.~(\ref{ModelH})-(\ref{ut}).
The position of the level $\varepsilon(t)$ as well as its coupling to the 
reservoir $u(t)$ are driven by a slowly changing external field, Eqs.~(\ref{et})-(\ref{gt}).

After gradient expansion,  
\begin{align}
&G^{r}(t_1,t_2) \to G^r(t,E) \\
&G^{<}(t_1,t_2) \to G^<(t,E)=i\, A(t,E)\,\phi(t,E),
\end{align}
where the system spectral function is given by 
\begin{equation}
A(t,E) \equiv -2\, \mbox{Im}\, G^r(t,E)
\end{equation}
and $\phi(t,E)$ is the non-equilibrium population of the level. 
Also
\begin{align}
&\Sigma^{r}(t_1,t_2) \to \Sigma^r(t,E) = \Lambda(t,E)-i\Gamma(t,E)/2 \\
&\Sigma^{<}(t_1,t_2) \to \Sigma^<(t,E) = i\, \Gamma(t,E)\, f(E),
\end{align}
where $\Lambda$ and $\Gamma$ are the Lamb shift and the broadening caused by the 
coupling to the reservoir and $f(E)$ is the Fermi-Dirac thermal distribution.  

We now apply the second order gradient expansion (\ref{grad2})
to expressions for the fluxes, Eqs.~(\ref{InuGF}) and (\ref{JnuGF}). 
This leads to
\begin{align}
\label{grad2I}
& I(t) = \int\frac{dE}{2\pi}\, I(t,E) = \frac{d}{dt}\int\frac{dE}{2\pi}A(t,E)\,\phi(t,E) \\
\label{grad2J}
& J_\alpha(t) = (\alpha-1)\,\partial_t\langle\hat V(t)\rangle
+\bigg(\frac{1}{2}-\alpha\bigg)\,d_t\langle \hat V(t)\rangle \\  
&\hspace{1.2cm}+ \int\frac{dE}{2\pi}\,E\,I(t,E),
\nonumber
\end{align}
where 
\begin{align}
\label{ItE}
&I(t,E)= \left\{E-\varepsilon(t); A\phi\right\} \\
\label{grad1V}
&\langle \hat V(t)\rangle = 2\int\frac{dE}{2\pi}\big(A\phi\,\Lambda 
+ \mbox{Re}\,G^r\,\Gamma f\big)\\
&\hspace{1.4cm}+ \frac{1}{2}\int\frac{dE}{2\pi}\big( \left\{\Gamma;A\phi\right\} 
-\left\{\Gamma\phi;A\right\} \big) \nonumber \\
\label{gard2ptV}
&\partial_t\langle\hat V(t)\rangle = \int\frac{dE}{2\pi}\big(A\phi\,\partial_t \Lambda 
+ \mbox{Re}\,G^r\,\partial_t \Gamma\,\phi \big) \\
&\hspace{1.6cm}+\frac{1}{4}\int\frac{dE}{2\pi}\big( \left\{\partial_t\Gamma;A\phi\right\} 
- \left\{\partial_t\Gamma\,\phi;A\right\} \big). \nonumber
\end{align} 
Note that evaluation of expressions (\ref{grad2I}) and (\ref{grad2J}) up to second order
in gradient expansion requires the knowledge of the $G^r$, $A$, and $\phi$ only up to 
first order (see Eqs.~(\ref{Gr0})-(\ref{phi1}) below).
Note also that in the spirit of the Botermans and Malfliet (BM) 
approximation~\cite{BotermansMalflietPhysRep90}, we substituted $f(E)$ by $\phi(t,E)$ 
in all the expressions involving derivatives of the lesser projection of the self-energy. 

The retarded projection of the Green function $G^r(t,E)$, the spectral function 
$A(t,E)$ and the non-equilibrium distribution $\phi(t,E)$ can be expanded as
\begin{align}
\label{DdefGr}
&G^r(t,E) = G^{r(0)}(t,E) + G^{r(1)}(t,E) + G^{r(2)}(t,E) + \ldots \\
\label{DdefA}
&A(t,E)= A^{(0)}(t,E) + A^{(1)}(t,E) + A^{(2)}(t,E) + \ldots \\
\label{Ddefphi}
&\phi(t,E) = \phi^{(0)}(t,E) +\phi^{(1)}(t,E) + \phi^{(2)}(t,E) + \ldots,
\end{align}
where the orders coincide with the orders of the gradient expansion.
Inserting this expansion in the gradient expansion expression for the 
Green function equations-of-motion (\ref{LEOM}) and (\ref{REOM}),
and identifying terms order by order, one finds 
that~\cite{IvanovKnollVoskresenskyNuclPhysA00,KitaProgrTheorPhys10},
\begin{align}
\label{Gr0}
&G^{r(0)}(t,E)=\left[E-\varepsilon(t)-\Sigma^r(t,E)\right]^{-1}
\\ 
\label{A0}
&A^{(0)}(t,E)= 
\frac{\Gamma(t,E)}{\left(E-\varepsilon(t)-\Lambda(t,E)\right)^2 + \left(\Gamma(t,E)/2\right)^2} 
\\
\label{phi0}
&\phi^{(0)}(t,E)=f(E) 
\end{align}
and 
\begin{align}
\label{Gr1A1}
&G^{r(1)}(t,E)=A^{(1)}(t,E)=0
\\
\label{phi1}
&\phi^{(1)}= -d_E f\,\frac{A^{(0)}}{2}
\bigg(d_t\varepsilon+\partial_t \Lambda+\partial_t\Gamma\,\frac{E-\varepsilon-\Lambda}{\Gamma}\bigg).
\end{align}

\subsection{Quasi-static driving}
The reversible transformation in the system is performed by a quasi-static driving, which 
corresponds to expanding the fluxes to first order in Eqs. ~(\ref{grad2I}) and (\ref{grad2J}).
To do so we only need the zero order correction of the retarded Green function 
$G^{r\,(0)}(t,E)$, its corresponding $A^{(0)}(t,E)$, and of the population $\phi^{(0)}(t,E)$.
We find
\begin{align}
\label{I1}
& I^{(1)}(t)= \int\frac{dE}{2\pi}\partial_t A^{(0)}\, f \\
& J^{(1)}_\alpha(t) = (\alpha-1)\big[\partial_t\langle\hat V(t)\rangle\big]^{(1)}
+\bigg(\frac{1}{2}-\alpha\bigg)d_t\langle\hat V(t)\rangle^{(0)} \nonumber \\ 
&\hspace{1.4cm} + \int\frac{dE}{2\pi}E\big(\partial_t A^{(0)}\, f 
+ d_t\varepsilon\,\partial_E\left(A\, f\right)\big) ,
\label{J1}
\end{align}
where
\begin{align}
\label{V0}
&\langle\hat V(t)\rangle^{(0)} =
2\int\frac{dE}{2\pi}\left( A^{(0)}f\,\Lambda +\mbox{Re}\, G^{r(0)}\,\Gamma f\right) \\
\label{ptV1}
& \big[\partial_t\langle\hat V(t)\rangle\big]^{(1)} =
\int\frac{dE}{2\pi}\left( A^{(0)}f\,\partial_t \Lambda 
+ \mbox{Re}\, G^{r(0)}\,\partial_t\Gamma \right).
\end{align}
Using (\ref{I1})-(\ref{ptV1}) in the definition (\ref{TradHeat}) yields Eq.~(\ref{QQS}).

Since both the Lamb shift, $\Lambda(t,E)$, and broadening, $\Gamma(t,E)$,  
are proportional to $u^2(t)$ (see Eqs.~(\ref{defLambda}) and (\ref{defGamma})),
and taking into account (\ref{QQS}), the condition (\ref{drv2}) means that the derivative of 
$\int dE\, f\, A^{(0)}\, d_t{\varepsilon}$ with respect to the driving parameter 
for the system-reservoir coupling $u(t)$ should be equal to the derivative of
$(1-\alpha) \int dE f \big[\mbox{Re}G^{r(0)}\,\partial_t{\Gamma}+A^{(0)}\,\partial_t{\Lambda}\big]$
with respect to the driving parameter for the level position $\varepsilon(t)$.
It is easy to see that this condition is satisfied only for $\alpha=0$.

Since the exact differential of the reversible heat defines entropy
\begin{equation}
T\,d_t S(t) = \dot{Q}_0^{(1)}(t),
\end{equation}
we find that the entropy is given (up to a constant) by
\begin{align}
\label{S1}
S =& \int\frac{dE}{2\pi}f\bigg(
A\bigg[\frac{E-\mu}{T}+\frac{E-\varepsilon}{T}\bigg] \\ &\qquad
+\frac{2}{T}\arctan\frac{E-\varepsilon-\Lambda}{\Gamma/2}\bigg).\nonumber
\end{align}
Utilizing 
\begin{align}
\frac{E-\mu}{T} =& \ln\frac{1-f(E)}{f(E)} \\
\frac{f(E)}{T} =& \frac{d}{dE}\ln[1-f(E)]
\end{align}
and performing an integration by parts for the last term in (\ref{S1}), we get
\begin{equation}
\begin{split}\label{EqEntropy}
S=& \int\frac{dE}{2\pi}A \big(-f\ln f-[1-f]\ln [1-f]\big)\\
&+ \int\frac{dE}{2\pi} A f \frac{(E-\varepsilon)}{T} \\
&+ \int\frac{dE}{2\pi} A \ln [1-f] \big(\partial_E \Lambda
 +\frac{E-\varepsilon-\Lambda}{\Gamma} \partial_E \Gamma \big).
\end{split}
\end{equation}
Note that in the limit of weak coupling, when $\Lambda\to 0$ and $\Gamma\to 0$, 
the entropy (\ref{EqEntropy}) reproduces the standard Shannon expression used 
in thermodynamics of weakly coupled systems. 

We stress that the quasi-static driving results do not rely on the BM approximation.

\subsection{Beyond quasi-static driving}

To calculate the fluxes (\ref{grad2I}) and (\ref{grad2J}) to second order, 
we therefore need corrections up to first order of the retarded Green 
function $G^{r\,(0,1)}(t,E)$, its corresponding $A^{(0,1)}(t,E)$, and of the 
nonequilibrium population $\phi^{(0,1)}(t,E)$. This leads to
\begin{align}
\label{I2}
&I^{(2)}(t)= 
\\ &\qquad
\int\frac{dE}{2\pi}\bigg(\partial_t \left(A^{(0)}\,\phi^{(1)}\right) 
+ d_t\varepsilon\,\partial_E\left(A^{(0)}\,\phi^{(1)}\right)\bigg)
\nonumber \\
\label{J2}
&J^{(2)}_\alpha(t) = 
\\ &\qquad
 (\alpha-1)\big[\partial_t\langle\hat V(t)\rangle\big]^{(2)}
+\bigg(\frac{1}{2}-\alpha\bigg)d_t\langle\hat V(t)\rangle^{(1)}
\nonumber \\ &\qquad +
\int\frac{dE}{2\pi}E\bigg(\partial_t \left(A^{(0)}\,\phi^{(1)}\right) 
+ d_t\varepsilon\,\partial_E\left(A^{(0)}\,\phi^{(1)}\right)\bigg),
\nonumber
\end{align}
where
\begin{align}
\label{V1}
&\langle\hat V(t)\rangle^{(1)} =
2\int\frac{dE}{2\pi}\,A^{(0)}\phi^{(1)}\,\Lambda 
 \\ &\qquad
+ \frac{1}{2}\int\frac{dE}{2\pi}\bigg( \left\{\Gamma;A^{(0)}\, f\right\}
-\left\{\Gamma\, f;A^{(0)}\right\}\bigg)
\nonumber \\
\label{ptV2}
&\big[\partial_t\langle\hat V(t)\rangle\big]^{(2)} =
\\ &\qquad
\int\frac{dE}{2\pi}\big(A^{(0)}\phi^{(1)}\,\partial_t \Lambda+
\mbox{Re}\,G^{r(0)}\,\partial_t \Gamma\,\phi^{(1)}
\big)
\nonumber \\ &\qquad
+\frac{1}{4}\int\frac{dE}{2\pi}\bigg(
\left\{\partial_t\Gamma;A^{(0)}\, f\right\} - \left\{\partial_t\Gamma\, f;A\right\}
\bigg).
\nonumber
\end{align}
Using (\ref{I2})-(\ref{ptV2}) in the definition (\ref{TradHeat}) yields
\begin{align}
\label{Q2}
&\dot{Q}_{\alpha}^{(2)} = \left(\frac{1}{2}-\alpha\right)\, d_t\langle\hat V(t)\rangle^{(1)}
\\ & +
\frac{d}{dt} \bigg( \int\frac{dE}{2\pi}A^{(0)}\bigg((E-\mu) \phi^{(1)}
+\frac{1-\alpha}{4}\partial_t\Gamma\partial_E f \bigg) 
\nonumber \\ 
&+\int\frac{dE}{2\pi}\partial_E f \frac{\left[A^{(0)}\right]^2}{2}
\bigg(d_t\varepsilon+\partial_t\Lambda+\partial_t\Gamma\frac{E-\varepsilon-\Lambda}{\Gamma}\bigg)
\nonumber \\ &\qquad\times
\bigg(d_t\varepsilon+(1-\alpha)\left(\partial_t\Lambda+\partial_t\Gamma\frac{E-\varepsilon-\Lambda}{\Gamma}\right)\bigg).
\nonumber
\end{align}
When considering periodic transformations where the system has reached a stationary regime,
the second law of thermodynamics states that 
\begin{equation}
Q_{0} = Q_{0}^{(2)} \leq 0,
\end{equation} 
where we used the fact that $\Delta S^{eq}=Q_{0}^{(1)}/T=0$.
We verify that this relation is satisfied since along such cyclic transformation 
only the last two lines of Eq.~(\ref{Q2}) survive and one finds that for $\alpha=0$ they become
\begin{equation}
\int\frac{dE}{2\pi}\partial_E f \frac{\left[A^{(0)}\right]^2}{2}
\bigg(d_t\varepsilon+\partial_t\Lambda+\partial_t\Gamma\frac{E-\varepsilon-\Lambda}{\Gamma}\bigg)^2
\end{equation}
which is indeed always negative or zero.

We now consider the wide band approximation (WBA) (i.e. $\Lambda=0$ and $\Gamma(E)=\Gamma=const$) 
and driving only in the level position and not in the coupling ($u(t)=1$) to show that the 
expressions (\ref{I1})-(\ref{QQS}) and (\ref{I2})-(\ref{Q2}) reduce to the results derived 
in Ref.~\cite{SanchezPRB14}. In this case we can make use of the identity
\begin{equation}
\label{trick}
\partial_t A^{(0)}(t,E) = -d_t\varepsilon(t)\,\partial_E A^{(0)}(t,E).
\end{equation}

We start by considering the particle current. 
Utilizing (\ref{trick}) in (\ref{I1}) and integrating by parts in energy leads to
\begin{equation}
\label{I1Sanchez}
 I^{(1)}(t) = d_t\varepsilon \int\frac{dE}{2\pi}\, d_E f\, A^{(0)}
\end{equation}
Similarly, utilizing (\ref{trick}) in (\ref{I2}) and integrating by parts in energy leads to
\begin{align}
\label{I2Sanchez}
I^{(2)}(t) =&  \partial_t\int\frac{dE}{2\pi}A^{(0)}\,\phi^{(1)}
\nonumber \\
\equiv& -\frac{1}{2}\partial_t\int\frac{dE}{2\pi}d_E f\,
\left[A^{(0)}\right]^2\,d_t\varepsilon ,
\end{align}
where the second equality is obtained by using the WBA version of (\ref{phi1}).
Expressions (\ref{I1Sanchez}) and (\ref{I2Sanchez}) are the results presented 
in equation (S.33) of the supporting information of Ref.~\cite{SanchezPRB14}. 
Note that difference in sign is due to our flux definition (positive when going 
from the reservoir to the system) which is opposite to the choice in Ref.~\cite{SanchezPRB14}.

We now turn to evaluating the coupling term. Using (\ref{V0}) within the WBA one gets
\begin{equation}
\left[d_t\langle\hat V(t)\rangle\right]^{(1)} \equiv d_t\langle\hat V(t)\rangle^{(0)}
= 2\int \frac{dE}{2\pi}\,\Gamma\, f\, \partial_t\mbox{Re}\, G^{r(0)}.
\end{equation}
Utilizing 
\begin{equation}
 d_t\mbox{Re}\, G^{r(0)} = -d_t\varepsilon\,\partial_E\mbox{Re}\, G^{r(0)}
\end{equation}  
and integrating in energy by parts leads to
\begin{align}
\label{dV1Sanchez}
\left[d_t\langle\hat V(t)\rangle\right]^{(1)} =& 
2\int\frac{dE}{2\pi}\, d_E f\,\Gamma\,\mbox{Re}\, G^{r(0)}\, d_t\varepsilon
\nonumber \\
\equiv&
2\int\frac{dE}{2\pi}\, d_E f\,A^{(0)}\,(E-\varepsilon)\, d_t\varepsilon.
\end{align} 
Similarly in the WBA (\ref{V1}) becomes
\begin{equation}
\left[d_t\langle\hat V(t)\rangle\right]^{(2)} \equiv d_t\langle\hat V(t)\rangle^{(1)}
= -\frac{1}{2}d_t \int \frac{dE}{2\pi}\,\Gamma\, d_E f\, \partial_t A^{(0)}.
\end{equation}
Since using (\ref{trick}),
\begin{equation}
\partial_t A^{(0)} = \frac{2\, (E-\varepsilon)\, d_t\varepsilon}{\Gamma}\left[A^{(0)}\right]^2,
\end{equation}
we get that
\begin{equation}
\label{dV2Sanchez}
\left[d_t\langle\hat V(t)\rangle\right]^{(2)} = -d_t\int\frac{dE}{2\pi}\, d_E f\,
\left[A^{(0)}\right]^2\, (E-\varepsilon)\, d_t\varepsilon.
\end{equation}
Expressions (\ref{dV1Sanchez}) and (\ref{dV2Sanchez}) are the results presented 
in equation (S.36) of the supporting information of Ref.~\cite{SanchezPRB14}. 

We finally turn to the energy current. Taking the choice $\alpha=1/2$ and 
disregarding the driving in the system-reservoir coupling (the first term) 
in Eq.~(\ref{J1}), after using (\ref{trick}), we get 
\begin{equation}
\label{J1Sanchez}
J_{1/2}^{(1)}(t) = \int\frac{dE}{2\pi}\, d_E f\, E\, A^{(0)}\, d_t\varepsilon .
\end{equation}
Similarly Eq.~(\ref{J2}) after employing (\ref{trick}) yields
\begin{equation}
J_{1/2}^{(2)}(t) = \int\frac{dE}{2\pi}\, E\, A^{(0)}
\left(\partial_t\phi^{(1)}+d_t\varepsilon\,\partial_E\phi^{(1)}\right).
\end{equation}
Substituting the WBA version of Eq.~(\ref{phi1}) and performing the derivatives leads to
\begin{align}
\label{J2Sanchez}
J_{1/2}^{(2)}(t) =& -\frac{1}{2}\int\frac{dE}{2\pi}\, d_E f
\\ &\times
\bigg( E\, d_t\left(\left[A^{(0)}\right]^2\, d_t\varepsilon\right)
-\left(A^{(0)}\, d_t\varepsilon\right)^2 \bigg).
\nonumber
\end{align}
Expressions (\ref{J1Sanchez}) and (\ref{J2Sanchez}) are the results presented 
in equation (S.32) of the supporting information of  Ref.~\cite{SanchezPRB14}. 
Once more, the difference in sign is due to our opposite convention for 
the flux compared to Ref.~\cite{SanchezPRB14}. 

\begin{acknowledgments}
M.E. is supported by the National Research Fund, Luxembourg in the frame of project FNR/A11/02.
M.G. gratefully acknowledges support by the Department of Energy (Early Career Award, DE-SC0006422).
\end{acknowledgments}


\end{document}